\documentclass[twocolumn,superscriptaddress,nobibnotes,aps,prd,showpacs,nofootinbib]{revtex4}
\usepackage{amsmath}
\usepackage{amsfonts}
\usepackage{amssymb}
\usepackage{graphicx}

\def\be{\begin{equation}}
\def\ee{\end{equation}}
\def\bea{\begin{eqnarray}}
\def\eea{\end{eqnarray}}

\renewcommand{\d}{\mathrm{d}}

\begin{document}

\title{Gravitational, lensing and stability properties of Bose-Einstein
condensate dark matter halos}
\author{Tiberiu Harko}
\email{t.harko@ucl.ac.uk}
\affiliation{Department of Mathematics, University College London, Gower Street, London
WC1E 6BT, United Kingdom}
\author{Francisco S. N. Lobo}
\email{fslobo@fc.ul.pt}
\affiliation{Instituto de Astrof\'{\i}sica e Ci\^{e}ncias do Espa\c{c}o, Faculdade de
Ci\^encias da Universidade de Lisboa, Edif\'{\i}cio C8, Campo Grande,
P-1749-016 Lisbon, Portugal}

\begin{abstract}
The possibility that dark matter, whose existence is inferred from the study
of the galactic rotation curves, and from the mass deficit in galaxy
clusters, can be in a form of a Bose-Einstein Condensate, has been
extensively investigated lately. In the present work, we consider a detailed
analysis of the astrophysical properties of the Bose-Einstein Condensate
dark matter halos that could provide clear observational signatures that 
help discriminate between different dark matter models. In the
Bose-Einstein condensation model dark matter can be described as a
non-relativistic, gravitationally confined Newtonian gas, whose density and
pressure are related by a polytropic equation of state with index $n=1$. The
mass and gravitational properties of the condensate halos are obtained in a
systematic form, including the mean logarithmic slopes of the density and of
the tangential velocity. The lensing properties of the condensate dark
matter are investigated in detail. In particular, a general analytical
formula for the surface density, an important quantity that defines the
lensing properties of a dark matter halos, is obtained in the form of series
expansions. This enables arbitrary-precision calculations of the surface
mass density, deflection angle, deflection potential, and of the
magnification factor, thus giving the possibility of the comparison of the
predicted lensing properties of the condensate dark matter halos with
observations. The stability properties of the condensate halos are also
investigated by using the scalar and the tensor virial theorems,
respectively, and the virial perturbation equation for condensate dark
matter halos is derived. As an application of the scalar virial theorem we consider the problem of the stability of a slowly rotating and slightly disturbed galactic dark matter halo. For such a halo the oscillations frequencies and the stability conditions are obtained in the linear approximation.
\end{abstract}

\date{\today}
\maketitle








\section{Introduction}


The polytropic gas model, in which the pressure $P$ has a simple power law
dependence of the matter energy density $\rho $, given by $P=K\rho ^{\gamma
} $, plays an important role in the modelling of self-gravitating systems of
compact astrophysical objects, such as white dwarfs and neutron stars, as
well as in the study of the equation of state of nuclear matter. Usually the
polytropic equation of state is represented as $P=K\rho ^{1+1/n}$, where $n$
is called the polytropic index. The Poisson and the hydrostatic equilibrium
equations for a self-gravitating, spherically symmetric polytropic fluid can
be combined to provide the basic equation describing the equilibrium
properties of the system as
\begin{equation}
\frac{1}{\zeta ^{2}}\frac{d}{d\zeta }\left( \zeta ^{2}\frac{d\theta }{d\zeta
}\right) +\theta ^{n}=0,  \label{1}
\end{equation}
where the dimensionless quantities $\theta $ and $\xi $ are defined as $\rho
=\rho _{c}\theta ^{n}$ and $\zeta =\sqrt{4\pi G\rho _{c}^{2}/\left(
n+1\right) P_{c}}r$, with $\rho _{c}$ and $P_{c}$ the central energy density
and pressure, respectively. Eq.~(\ref{1}) is known as the Lane-Emden
equation, and its physical, astrophysical and mathematical properties have
been intensively studied by using both analytical and numerical methods. It
admits only three exact solutions, corresponding to $n=0$, $n=1$ and $n=3$,
respectively \cite{Hor}.

From these three solutions, the unusual solution corresponding to $n=1$,
having the form $\theta =\sin \zeta /\zeta $, does not seem at first sight
to have many realistic physical properties, and consequently it has been
less investigated, as compared to the other solutions. However, somehow
unexpectedly, the $n=1$ polytropic equation of state appears in a
fundamental field of research apparently unrelated to astrophysics, namely,
the theory of the condensed Bose-Einstein cold atomic gases. A basic result
in quantum statistical mechanics is that at very low temperatures, in a
dilute Bosonic system (Bose gas) all particles condense to the same quantum
ground state, forming a so called Bose-Einstein condensate (BEC). From a
physical point of view a BEC is characterized by a sharp peak in both
coordinate and momentum space distribution. The BEC phase transition takes
place when the particles are correlated with each other quantum,
mechanically. This happens when their wavelengths overlap, that is, the
thermal de Broglie wavelength $\lambda _{T} $ becomes greater than the mean
inter-particles distance $l$. Hence the BEC transition occurs at a
temperature
\begin{equation}  \label{temp}
T<\left(\frac{2\pi \hbar ^{2}}{mk_{B}}\right)n^{2/3},
\end{equation}
where $m$ is the mass of the particle in the condensate, $n$ is the number
density, and $k_{B}$ is Boltzmann's constant \cite{Da99}-\cite{rev7}. A
coherent quantum condensate state develops in two physical situations: a)
when the particle density is high enough, or b) the system temperature is
sufficiently low. The Bose-Einstein Condensation has been observed in
laboratory experiments \cite{exp1,exp2,exp3}. A large number of quantum
degenerate gases have been created by a combination of laser, magnetic and
evaporative cooling techniques. The experimental investigations have opened
several new directions of research, involving atomic, statistical and
condensed matter physics \cite{Da99}-\cite{exp3}.

The purpose of this paper will be applications of the BEC to the dark matter
problem. Indeed, the existence of dark matter is primarily supported by the
rotation curve of galaxies \cite{dm1, dm2,dm3}, and also by the difference
between the masses derived from luminosity and from the virial theorem in
clusters of galaxies \cite{Giov}. Furthermore, gravitational lensing data
also indicates that light deflection is more accentuated than would be to
visible matter alone \citep{Massey}. Thus, by introducing a new matter
component, denoted dark matter, it is possible to explain all of these
observational phenomena. However, the nature of dark matter, and its
constituent particle(s), still remain essentially unknown, despite several
decades of intensive research. For reviews on the dark matter problem see
\cite{Massey, OvWe04}.

In this context, the possibility that dark matter could be in the form of a
Bose-Einstein condensate was analyzed in detail in \cite{BoHa07} (for
earlier work in this field see \cite{Sin1,Sin2,Lee96,Fer}). In particular,
the equation of state of the condensate dark matter was obtained as a
polytropic equation of state of index $n=1$, and the corresponding density
profile was discussed. The theoretical rotation curves were fitted with the
observational data from a sample of Low Surface Brightness (LSB) galaxies,
and it was pointed out that the lensing properties may discriminate between
the standard pressureless and the condensate dark matter models. The
physical and astrophysical properties of Bose-Einstein Condensate dark
matter halos were investigated in \cite{inv}.

Recently, the rotation curves in the Bose-Einstein condensate dark matter
model were analyzed in \cite{Mat}, and a good agreement with observations
was found. A comparison of the predicted rotation curves with the
observational data for eight dwarf galaxies was performed in \cite{Har2},
and it was shown that the presence of the condensate dark matter can solve
the core/cusp problem faced by the standard $\Lambda $CDM ($\Lambda $Cold
Dark Matter) model. The effects of the angular momentum and on the vortices
on the equilibrium of self-gravitating, rotating BEC haloes which satisfy
the Gross-Pitaevskii-Poisson equations were considered in \cite{Brook,
Rind,Rind1,kai,Zin}. In a Bose-Einstein condensate vortices form as long as
the self-interaction is strong enough, which is not the case for axionic
dark matter. The galactic masses in the framework of the condensate model
were discussed in \cite{Lee09,Lee10}, and the effects of the finite
temperature of the condensate on the density profiles were studied in \cite%
{Har4}. The equilibrium properties of Newtonian self-gravitating
Bose-Einstein condensates with short-range interactions were investigated in
detail in \cite{Chav1,Chav2}. The study of the cosmological implications of
the Bose-Einstein condensation has also become an active field of research %
\citep{Fuk08, Fuk09, Har1,Chav3,Chav4,Har3,Har5,Wamba, kai1, NC}. In
particular, in \cite{Wamba} it was shown that condensate dark matter effects
can be seen in the CMB matter power spectrum if the mass $m_{\chi }$ of the
condensate particle lies in the range 15 meV $< m_{\chi } < 700$ meV,
leading to a small, but perceptible, excess of power at large scales.
The possibility of the existence of the BEC stars was also considered \cite%
{stars}, while BEC string models were investigated in \cite{string}.

One of the central issue in the study of Bose-Einstein Condensate dark
matter halos is the nature, and properties, of the dark matter particle. One
of the most interesting dark matter particle candidate is the axion, a
(still) hypothetical elementary particle postulated by the Peccei-Quinn
theory \cite{PQ} to resolve the strong CP problem in quantum chromodynamics.
It was shown in \cite{axion} that dark matter axions do form a BEC, as a
result either of their self interactions, or as a result of their
gravitational interactions. It was also proven, from the study of the
cosmological perturbations, that axion Bose-Einstein Condensate dark matter
halos differ from standard Cold Dark Matter ones on small scales only.
Unlike vortices in superfluid He$^4$ and dilute gases, the vortices in the
axion Bose-Einstein Condensate dark matter are attractive \cite{axion0}.
Therefore a large fraction of the vortices in the axion BEC may form a
single big vortex along the rotation axis of the galaxy. The resulting
enhancement of caustic rings could explain the rises in the Milky Way
rotation curve, usually attributed to caustic rings. The properties of the
axionic dark matter have been extensively investigated in \cite{axion1}.

It is the purpose of the present paper to extend the previous investigations
of the $n=1$ polytropic Bose-Einstein dark matter halos, by considering in
detail and in a systematic way the observationally relevant physical
parameters. As a first step in this study we consider the mass and
gravitational properties of the dark matter halos in the condensate model.
The gravitational potential and energy of the halo, its kinetic energy, as
well as the observationally extremely important logarithmic slopes of the
density and velocity profiles are obtained. One of the most important method
which could discriminate between the condensate model and other dark matter
models is gravitational lensing. We obtain and present in detail, in an
analytical form, the physical parameters (surface mass density, deflection
angle and magnification etc.) necessary for the analysis and comparison of
the observational data with the predictions of the model. Finally, the
formulations of the virial theorem are obtained for rotating and
non-rotating dark matter halos, the virial perturbation equation is derived,
and some general stability conditions are formulated. As an application of the scalar virial theorem we consider in detail the problem of the stability of a slowly rotating and slightly disturbed galactic dark matter halo. By introducing a mathematical description based on the Lagrangian displacements, from the scalar virial theorem we obtain the equation describing the time evolution of the perturbations. The oscillations frequencies and the stability conditions of the Bose-Einstein Condensate dark matter galactic halo are obtained in the linear approximation.

The present paper is organized as follows. In Section~\ref{sect2}, the basic
equations describing the Bose-Einstein condensate dark matter halos are
presented. In Section~\ref{sect3}, the general gravitational properties of
the condensate halos are investigated. The gravitational lensing by dark
matter halos is considered in Section~\ref{sect4}. The virial theorem for
Bose-Einstein Condensate dark matter halos and their stability properties
are presented in Section~\ref{sect5}. As an application of the scalar virial theorem we consider the problem of the stability of a slowly rotating and slightly disturbed galactic dark matter halo is considered in Section~\ref{6n}. We discuss and conclude our results in
Section~\ref{sect6}.


\section{Bose-Einstein condensate dark matter halos}

\label{sect2}

The transition of the dark bosonic component to Bose-Einstein Condensed
state is assumed to have taken place during the cosmological expansion of
the Universe at a redshift of around $z\approx 1400-1500$ \cite{Har5}. At
that moment the temperature and the matter density of the Universe did
satisfy the condition given by Eq.~(\ref{temp}). Due to the expansion of the
Universe, its temperature and density further decreased. In the following we
make the fundamental assumption that the dark matter halos are composed of a
strongly-coupled cold dilute Bose-Einstein condensate, at absolute zero
temperature. Hence this assumption implies that almost all of the dark
matter particles are in the condensate. Since the dark matter halo is dilute
and cold, only binary collisions at low energy between dark matter particles
are relevant. Therefore, independently of the details of the two-body
potential, it follows that the collisions can be characterized by a single
physical parameter, the $s$-wave scattering length $l_a$ \cite{Da99,rev6}.
Therefore, with an excellent approximation, one can substitute the
interaction potential with an effective interaction term of the form $%
V_I\left( \vec{r}^{\prime }-\vec{r}\right) =U_0 \delta \left( \vec{r}%
^{\prime }-\vec{r}\right) $, where the coupling constant $U_0 $ is related
to the scattering length $l_a$ through the relation $U_0 =4\pi \hbar
^{2}l_a/m_{\chi }$, where $m_{\chi}$ is the mass of the dark matter particle %
\citep{Da99,rev6}. In this approach the ground state properties of the dark
matter are described by the mean-field Gross-Pitaevskii (GP) equation \cite%
{Da99,rev6,rev7}. The GP equation for the dark matter halos can be derived
from the GP energy functional \cite{Da99,rev6,rev7, Lush},
\begin{eqnarray}
E\left[ \psi \right] &=&\int \Bigg[ \frac{\hbar ^{2}}{2m_{\chi }}\left|
\nabla \psi \left( \vec{r}\right) \right| ^{2}+ \frac{U_{0}}{2}\left| \psi
\left( \vec{r}\right) \right| ^{4}-\nonumber\\
&&m_{\chi}\psi ^{*}\left( \vec{r}\right) \vec{\Omega}\cdot\left(\vec{r}\times \vec{p}\right)\psi\left( \vec{r}\right) \Bigg] d^3\vec{r} - \notag \\
&&  \frac{1}{2}Gm_{\chi }^{2} \int \int \frac{\left| \psi \left( \vec{r}%
\right) \right| ^{2}\left| \psi \left( \vec{r}^{\prime }\right) \right| ^{2}%
}{\left| \vec{r}-\vec{r}^{\prime }\right| }d^3\vec{r}d^3\vec{r}^{\prime }
\notag \\
&=& E_K+E_{int}+E_{rot}+E_{grav},
\end{eqnarray}
where $\psi \left( \vec{r}\right) $ is the wave function of the condensate %
\citep{Da99,rev3,rev4,rev5,rev6,rev7}. The first term in the energy
functional is the quantum pressure (kinetic energy $E_K$), the second is the
interaction energy $E_{int}$, the third term is the rotational energy $E_{rot}$, with $\vec{\Omega} $ the angular velocity, and $\vec{p}$ the condensate momentum, and the fourth term is the gravitational potential
energy, respectively. The mass density of the condensate is defined as
\begin{equation}
\rho \left( \vec{r}\right) =m_{\chi}\left| \psi \left( \vec{r}\right)
\right| ^{2},
\end{equation}
and the normalization condition is $N=\int \left| \psi \left( \vec{r}\right)
\right| ^{2}d^3\vec{r}$, where $N$ is the total number of dark matter
particles. The variational procedure
\begin{equation}
\delta E\left[ \psi \right] -\mu \delta \int \left| \psi \left( \vec{r}%
\right) \right| ^{2}d^3\vec{r}=0,
\end{equation}
provides the GP equation as \citep{Da99,rev3,rev4, rev5,rev6,rev7}
\begin{eqnarray}  \label{GP}
-\frac{\hbar ^{2}}{2m}\nabla ^{2}\psi \left( \vec{r}\right)+m_{\chi}V_{grav}%
\left( \vec{r}\right) \psi \left( \vec{r}\right)  \notag \\
+U_{0}\left| \psi \left( \vec{r}\right) \right| ^{2}\psi \left( \vec{r}%
\right) -\Omega L_z\psi \left( \vec{r}\right)= \mu \psi \left( \vec{r}\right),
\end{eqnarray}
where the Lagrangian multiplier $\mu $ is the chemical potential, and the
confining gravitational potential $V_{grav}$ satisfies the Poisson equation
\begin{equation}  \label{poi}
\nabla ^{2}V_{grav}=4\pi G\rho .
\end{equation}

The term $−\Omega L_z = i\hbar\left(x\partial _y  y\partial _x\right)$ is due to the rotation of the system about the $z$ axis with an angular frequency $\Omega $. For a rotating Bose-Einstein Condensate dark matter halo, its rotational energy may be approximated as \cite{rev7}
\be
E_{rot}=\frac{1}{2}m_{\chi}\int_V{\omega ^2(\Omega)\left(x^2+y^2\right)\left |\psi (\vec{r})\right |^2d^3\vec{r}},
\ee
where $\omega $ is some function of the angular velocity of the rotation $\Omega $. 

In the physically important case when the number of particles in the
condensate becomes large enough, the quantum pressure term makes a
significant contribution to the thermodynamical parameters only near the
boundary, and it is much smaller than the interaction energy term. Thus, for
this case the quantum pressure term can be neglected (the Thomas-Fermi
approximation). When $N\rightarrow \infty $, it can be proven explicitly
that the Thomas-Fermi approximation becomes exact
\citep{Da99,rev3,rev4,
rev5, rev6, rev7}. Therefore we obtain
\begin{equation}\label{poin}
\rho \left( \vec{r}\right) =\frac{m_{\chi}}{U_{0}} \left[ \mu
-m_{\chi}V_{grav}\left( \vec{r}\right) -\frac{1}{2}m_{\chi}\omega ^2\left(x^2+y^2\right)\right].
\end{equation}

In the non-rotating case besides of the conservation of the total energy $E=E_0+\int{m_{\chi}V_{grav}%
\left|\psi \right|^2d^3\vec{r}}$, where
\begin{equation}
E_0=\int \left[ \frac{\hbar ^{2}}{2m_{\chi}}\left| \nabla \psi \left( \vec{r}%
\right) \right| ^{2}+\frac{U_{0}}{2}\left| \psi \left( \vec{r}\right)
\right| ^{4}\right] d^3\vec{r},
\end{equation}
the GP energy functional provides two more conservation laws, the continuity
equation
\begin{equation}  \label{cont1}
\frac{\partial \rho }{\partial t}+\nabla \cdot \vec{j}=0,
\end{equation}
where the momentum $\vec{j}$ is given by
\begin{equation}  \label{mom}
\vec{j}=\frac{i\hbar }{2}\left( \psi \nabla \psi ^{\ast }-\psi ^{\ast
}\nabla \psi \right) ,
\end{equation}
and the momentum conservation equation \citep{Pit,Sed},
\begin{equation}  \label{cont2}
\frac{\partial j_{i}}{\partial t}+\frac{\partial \Pi _{ik}}{\partial x_{k}}%
=-\rho \frac{\partial V_{grav}}{\partial x_{i}}, \qquad i,k=1,2,3,
\end{equation}
with
\begin{equation}
\Pi _{ik}=\frac{\hbar ^{2}}{4m_{\chi}}\left( \frac{\partial \psi }{\partial
x_{i}}\frac{\partial \psi ^{\ast }}{\partial x_{k}}-\psi \frac{\partial
^{2}\psi ^{\ast }}{\partial x_{i}\partial x_{k}}+\mathrm{c.c}\right)
+p\delta _{ik},
\end{equation}
where c.c denotes complex conjugation, and the quantum pressure $p$ is given
by
\begin{equation}  \label{pressn}
p=\frac{U_{0}}{2m_{\chi }} \rho ^{2}= \frac{2\pi \hbar ^{2}l_a}{m_{\chi }^{2}%
} \rho ^{2}.
\end{equation}

By representing the wave function as $\psi \left( \vec{r},t\right) =\rho
\left( \vec{r},t\right) \exp \left[ i\varphi \left( \vec{r},t\right) \right]
$, the velocity of the condensate can be defined as $v\left( \vec{r}%
,t\right) =\left( \hbar /m_{\chi }\right) \nabla \varphi \left( \vec{r}%
,t\right) $. Then Eq.~(\ref{cont1}) leads to the mass conservation equation
\begin{equation}
\frac{\partial \rho }{\partial t}+\nabla \cdot \left( \rho \vec{v}\right) =0,
\end{equation}
and taking into account the Thomas-Fermi approximation, Eq.~(\ref{cont2})
gives the basic evolution equation of the Bose-Einstein Condensate dark
matter halos,
\begin{equation}  \label{4}
\frac{\partial }{\partial t}\rho v_{i}+\frac{\partial }{\partial x_{k}}%
\left( \rho v_{i}v_{k}+p\delta _{ik}\right) =-\rho \frac{\partial V_{grav}}{%
\partial x_{i}}.
\end{equation}

Formally, Eq.~(\ref{4}) is identical to the Euler equation in standard
hydrodynamics, however, the distinctive feature of the Bose-Einstein
condensate is that its flow is in general irrotational, $\nabla \times \vec{v%
}=0$ \citep{Sed,rev7}. After applying the $\nabla ^2$ operator on both sides
of Eq.~(\ref{poin}), the Poisson equation becomes
\begin{equation}  \label{eqpoi}
\nabla ^{2}\rho +k^{2}\rho =0,
\end{equation}
where we have denoted
\begin{equation}
k=\sqrt{\frac{4\pi Gm_{\chi}^{2}}{U_{0}}}\,,
\end{equation}
for notational simplicity.

Note that if the condensate is rotating at an angular velocity $\vec{\Omega }
$ that is greater than the critical one $\vec{\Omega}_{cr}$, its energy is
minimized via a creation of vortices. Hence it follows that the phase of the
condensate order parameter changes by $2\pi $ around a path that includes
vortex lines,
\begin{equation}
\nabla \times \vec{v}=\frac{\hbar }{m_{\chi }}\nabla \times \nabla \varphi
\left( \vec{r}\right) =\frac{2\pi \hbar }{m_{\chi }}\vec{n}\sum_{j}\delta
^{(2)}\left( \vec{r}-\vec{r}_{j}\right) ,  \label{5}
\end{equation}
where $\vec{n}$ is a unitary vector along a vortex line, $\vec{r}_{j}$ is
the radius vector of a vortex line in the plane orthogonal to the unit
vector $\vec{n} $, and $\delta ^{(2)}$ is a two-dimensional delta function
in the corresponding plane \cite{Sed,rev7}.

In order to write the Euler equation Eq.~(\ref{4}) in a frame rotating
uniformly with the angular velocity $\vec{\Omega }$ we need to add to the
right-hand side of the Euler equation the centrifugal potential $\left| \vec{%
\Omega}\times \vec{r}\right| /2$, and the Coriolis acceleration $2\vec{u}%
\times \vec{\Omega}$, respectively, where $\vec{u}$ is the condensate
velocity in the rotating frame. Therefore the equations of motion of an
inviscid Bose-Einstein Condensate, rotating at a constant angular velocity $%
\vec{\Omega}$ in the presence of a confining gravitational field are given,
in Cartesian coordinates, by \citep{Sed,rev7}
\begin{equation}
\rho \frac{du_{j}}{dt}-2\rho \varepsilon _{jkl}\Omega _{k}u_{l}=-\frac{%
\partial p}{\partial x_{j}}-\rho \frac{\partial }{\partial x_{j}}\left(
V_{grav}-\frac{1}{2}\left| \vec{\Omega}\times \vec{r}\right| ^{2}\right) ,
\label{6}
\end{equation}
where $\varepsilon _{jkl}$ is the completely antisymmetric unit tensor of
rank three.


\section{Mass and gravitational properties of static condensed dark matter
halos}

\label{sect3}

For static (non-rotating) condensates $\omega =0$, the general solution of
Eq.~(\ref{eqpoi}), describing the density distribution $\rho $ of the static
gravitationally bounded single component dark matter Bose-Einstein
condensate is given by \citep{BoHa07}
\begin{equation}  \label{neq1}
\rho \left( r\right) =\rho _{c}\frac{\sin kr}{kr}
\end{equation}
where $\rho _{c}$ is the central density of the condensate, $\rho _{c}=\rho
(0)$. Therefore the Bose-Einstein condensate dark matter profile is given by
the $n=1$ polytropic density profile.

The $n=1$ polytropic Bose-Einstein condensate dark matter density profile
has a well defined boundary radius $R$, at which the dark matter density
vanishes, so that $\rho (R)=0$. From this condition we obtain $R$ in the
form $kR=\pi $, which fixes the radius of the condensate dark matter halo as
\begin{equation}  \label{rad}
R=\frac{\pi}{k}=\pi \sqrt{\frac{\hbar ^{2}l_{a}}{Gm_{\chi}^{3}}}.
\end{equation}

For $r>R$ the density is smaller than zero, so that the $n=1$ polytropic
density profile cannot be extended to infinity, that is, beyond its sharp
boundary. For small values of $r$ the condensate dark matter profile can be
written as
\begin{equation}
\rho =\rho _c\left(1 - \frac{{\pi }^2}{6R^2}r^2+ \frac{{\pi }^4}{120R^4}%
r^4+... \right).
\end{equation}
The logarithmic slope $\alpha _{BE}$ of the Bose-Einstein profile is given
by \cite{Oh}
\begin{equation}  \label{ds}
\alpha _{DM} (r)=-\frac{d\ln \rho }{d\ln r}=1-kr\cot \left( kr\right) .
\end{equation}
Note that Eq.~(\ref{ds}) is not of the power law form $\alpha _{DM} \left(
r\right) \sim r^{n }$, where $n $ is a positive number defining the
steepness of the power law, as are most of the density profiles used in the
current astrophysical research. For $r\rightarrow 0$, $\alpha _{DM}(0)=0$,
while at the surface of the dark halo, where $kR=\pi $, $\alpha _{DM} $
diverges, so that $\lim_{kr\rightarrow \pi }\alpha _{DM}=-\infty $. For
small values of $r$ the logarithmic slope can be written as $\alpha _{DM}
(r)\approx k^{2}r^{2}/3+k^{4}r^{4}/45+O(r)^6$.

The mass profile $m(r)=4\pi \int_{0}^{r}\rho (r)r^{2}dr$ of the
Bose-Einstein condensate galactic halo is
\begin{eqnarray}
m\left( r\right) &=&\frac{4\pi \rho _{c}}{k^{2}}r\left[ \frac{\sin \left(
kr\right) }{kr}-\cos \left( kr\right) \right]  \notag \\
&=&\frac{4}{\pi}R^2r\rho (r)\alpha _{DM} (r).
\end{eqnarray}
The total mass $M$ of the condensate is
\begin{equation}  \label{26}
M(R)=\frac{4\pi ^2 \rho _{c}}{k^{3}}=\frac{4}{\pi }\rho _{c}R^{3},
\end{equation}
which represents a simple cubic proportionality between mass and radius. The
mass of the condensate is around three times smaller than the mass of a
constant $\rho _c$ density sphere. The central density of the condensate is
determined by the total mass $M(R)$ and the radius of the condensate as
\begin{equation}
\rho _c=\frac{\pi M(R)}{4R^3}.
\end{equation}

The mean density $\left\langle \rho \right\rangle =3M/4\pi R^{3}$ of the
condensate can be obtained as
\begin{equation}
\left\langle \rho \right\rangle =\frac{3\rho _{c}}{k^{2}R^{2}}=\frac{3\rho
_{c}}{\pi ^{2}}.
\end{equation}
Alternatively, we can define the mean density of the dark matter halo as
\begin{equation}
\bar{\rho }=\frac{1}{R}\int_0^R{\rho (r)dr}=\frac{\mathrm{Si}(\pi )}{\pi}%
\rho _c=0.5895\rho _c,
\end{equation}
where $\mathrm{Si}(x)=\int_0^x{\sin (x)dx/x}$, and $\mathrm{Si}(\pi)=1.8519$.

The gravitational potential $V_{grav} \left( r\right) $ of the condensed
dark matter distribution is determined by the equation
\begin{eqnarray}
V_{grav} (r)&=&G\int_{r}^{R}\frac{m\left( r^{\prime }\right) dr^{\prime }}{%
r^{\prime 2}}  \notag \\
&=&\frac{4G\rho _cR^3}{\pi ^2r} \sin \left(\frac{\pi r}{R}\right), \qquad
r\leq R.
\end{eqnarray}
At small radii the potential behaves as
\begin{equation}
V_{grav} (r)\approx \frac{4G\rho _cR^2}{\pi } - \frac{2\pi G \rho _c}{3}r^2
+ \frac{\pi ^3G\rho _c}{30R^2}r^4 + O(r)^6,
\end{equation}
for $r\leq R$.

The gravitational potential energy $U(r)$ per unit mass and inside radius $r$
of the condensed dark matter halo is given by
\begin{eqnarray}
U(r)&=&-4\pi G\int_{0}^{r}\frac{\rho (r)m(r)}{r}r^{2}dr=-\frac{2GR^4\rho _c^2%
}{\pi^3}\times  \notag \\
&& \left\{ 2\pi r \left[ 2 + \cos \left(\frac{2\pi r}{R}\right) \right] -
3R\sin \left(\frac{2\pi r}{R}\right) \right\}.
\end{eqnarray}
The total potential energy of the halo is given by
\begin{equation}
U(R)=-\frac{12\pi ^{3}G}{k^{5}}\rho _{c}^{2}=-\frac{12G\rho _c ^2}{\pi ^2}%
R^5.
\end{equation}

The intrinsic velocity dispersion for an isotropic model can be obtained
from the definition
\begin{equation}
\langle v_r^2(r)\rangle=\frac{1}{\rho (r)}\int_r^R{\frac{GM\left(r^{\prime
}\right)\rho \left(r^{\prime }\right)dr^{\prime }}{r^{\prime 2}}},
\end{equation}
and is given by
\begin{equation}
\langle v_r^2(r)\rangle=\frac{2 G R^3 \rho _c }{\pi ^2 r}\sin \left(\frac{%
\pi r}{R}\right)=\frac{2 G R^2 }{\pi }\rho (r).
\end{equation}

One can define a kinetic energy $K$ of the halo in terms of the average
velocity dispersion $\sigma _{V}$ as
\begin{equation}
K(r)=\frac{3}{2}\int \rho \left( r\right) \sigma _{V}^{2}(r)dV.
\end{equation}

If the velocity dispersion is a constant, we obtain
\begin{equation}
K(r)=\frac{6\pi \rho _{c}}{k^{2}}\sigma _{V}^{2}r\left[ \frac{\sin \left(
kr\right) }{kr}-\cos \left( kr\right) \right] ,
\end{equation}
and
\begin{equation}
K(R)=\frac{6\pi ^{2}\rho _{c}}{k^{3}}\sigma _{V}^{2}=\frac{6\rho _c}{\pi}%
\sigma _V^2R^3.
\end{equation}
For the ratio of the kinetic and potential energy, we obtain
\begin{equation}
\frac{K(R)}{\left| U(R)\right| }=\frac{k^{2}}{2\pi G\rho _{c}}\sigma
_{V}^{2}=\frac{\pi }{2G\rho _cR^2}\sigma _V^2.
\end{equation}

The tangential velocity of a test particle moving in the condensed dark halo
can be represented as \citep{BoHa07}
\begin{eqnarray}
V^{2}\left( r\right) &=&\frac{Gm(r)}{r}=\frac{4\pi G\rho _{c}}{k^{2}}\left[
\frac{\sin \left( kr\right) }{kr}-\cos \left( kr\right) \right]  \notag \\
&=&\frac{4GR^2}{\pi }\rho (r)\alpha _{DM} (r) .
\end{eqnarray}
The velocity at the vacuum boundary of the halo has the maximum value
\begin{equation}
V^{2}\left( R\right) =\frac{4\pi G\rho _{c}}{k^{2}}=\frac{4G\rho _cR^2}{\pi }%
=\frac{GM(R)}{R}.
\end{equation}
Therefore
\begin{equation}
\frac{K(R)}{\left| U(R)\right| }=2\frac{\sigma _{V}^{2}}{V^{2}\left(
R\right) }=\frac{2R}{GM(R)}\sigma _V^2.
\end{equation}

Similarly to the case of the density we can define a mean value of the
tangential velocity as $\bar{V^2}=(1/R)\int_0^R{V^2(r)dr}$, and which is
given by
\begin{equation}
\bar{V^2}=\frac{4G}{\pi ^2}\mathrm{Si}(\pi )\rho _c R^2.
\end{equation}
The logarithmic slope of the tangential velocity is defined as $\beta
_V=-d\ln V/d\ln r$, and is given by
\begin{eqnarray}
\beta_V&=&\frac{1}{2}\left[1 +\frac{1}{R} \frac{{\pi }^2r^2} { \pi r\cot
(\pi r/R)-R}\right]  \notag \\
&=&\frac{1}{2}\left[1-\frac{\pi ^2}{\alpha _{DM} (r)}\left(\frac{r}{R}%
\right)^2\right].
\end{eqnarray}
When $r=0$, $\beta _V=-1$, while $\beta _V=1/2 $ at the vacuum boundary of
the dark matter halo.

One can define the core (inner) radius $R_{core}$ of the Bose-Einstein
condensate dark matter halos as the radius which satisfies the condition $%
\alpha _{DM}\left(R_{core}\right)=1$ \citep{Mat}. By taking into account the
definition of the logarithmic density slope, we obtain for the core radius
the relation $kR_{core}=\pi/2$, or
\begin{equation}
R_{core}=\frac{R}{2}.
\end{equation}

The mean value of the logarithmic density slope within the radius $0\leq
r\leq R_{core}$ can be defined as \cite{Oh}
\begin{eqnarray}
\langle\alpha _{DM}\rangle &=& \frac{1}{R_{core}}\int_0^{R_{core}}{\alpha
_{DM}(r)}  \notag \\
&=& 1-\frac{2}{\pi}\int_0^{\pi /2}{x\cot x dx},
\end{eqnarray}
and it has the universal value
\begin{equation}  \label{slope1}
\langle\alpha _{DM}\rangle=0.3068.
\end{equation}

For the mean of the logarithmic slope of the tangential velocity we obtain
\begin{eqnarray}
\langle \beta _V\rangle &=&\frac{1}{R_{core}}\int _0^{R_{core}}{\beta _V
(r)dr}  \notag \\
&=&\frac{1}{R}\int_0^{R/2}{\left[1-\pi^2\left(\frac{r}{R}\right)^2\right]}dr
\notag \\
&=&\frac{1}{2}\left(1-\frac{\pi^2}{12}\right)=0.088.
\end{eqnarray}

The density corresponding to the inner radius of the condensed dark matter
halo is
\begin{equation}
\rho \left(R_{core}\right)=\frac{2}{\pi }\rho _c.
\end{equation}
Since the radius $R$ (and consequently also the inner core boundary $%
R_{core} $) of the dark matter halo is a universal constant, depending only
on the fundamental physical constants and the mass and scattering length of
the dark matter particles, it follows that for a given central density the
product $\rho _cR_{core}$ is a universal constant,
\begin{equation}
\rho _cR_{core}=\frac{\pi}{8}\frac{M(R)}{R^2}=\mathrm{constant}.
\end{equation}


\section{Lensing properties of condensed dark matter halos}

\label{sect4}

One of the important ways one could test the Bose-Einstein condensate
galactic dark matter model is by studying the light deflection by galaxies,
and in particular by studying the deflection of photons passing through the
region where the rotation curves are flat. In the present Section, we
consider the basic lensing properties of the condensate dark matter halos.
For a discussion of the gravitational lens properties of scalar field dark
matter haloes see \cite{Mielke1}.


\subsection{Surface mass density}


The surface mass density of a spherically symmetric lens is obtained by
integrating along the line of sight of the three-dimensional density
profile, $\Sigma \left( \xi \right) =\int_{-\infty }^{+\infty }\rho \left(
\xi ,r\right) \d z$, where $\xi $ is the radius measured from the center of
the lens and $r=\sqrt{\xi ^{2}+z^{2}}$. The surface mass density can be
written as an Abel transform \cite{dm3}, so that
\begin{equation}
\Sigma \left( \xi \right) =2\int_{\xi }^{+\infty }\frac{\rho \left( r\right)
rdr}{\sqrt{r^{2}-\xi ^{2}}}.
\end{equation}
By inserting the density profile of the Bose-Einstein condensate dark matter
halos we find
\begin{equation}
\Sigma \left( \xi \right) =2\rho _{c}\frac{R}{\pi} \int_{\xi }^{R}\frac{\sin
\left( kr\right) dr}{\sqrt{r^{2}-\xi ^{2}}}.
\end{equation}
Now, introducing a new variable $\vartheta $ by means of the transformation $%
r=\xi \cosh \vartheta $, we obtain
\begin{equation}
\Sigma \left( \xi \right) =2\rho _{c} {\frac{R }{\pi}} \int_{0}^{\mathrm{%
arccosh}\left( R/\xi \right) }\sin \left[ k\xi \cosh \vartheta \right]
d\vartheta .
\end{equation}

By taking into account the expansion
\begin{eqnarray}
\sin \left( k\xi \cosh\vartheta \right) =\sum_{s=0}^{\infty }\frac{\left(
-1\right) ^{s}2^{-\left( 2s+1\right) }}{\left( 2s+1\right) !}k^{2s+1}\xi
^{2s+1}  \notag \\
\times \sum_{l=0}^{2s+1}C_{l}^{2s+1}e^{\left( 2s-2l+1\right) \vartheta },
\end{eqnarray}%
where $C_{l}^{s}=s!/l!\left( s-l\right) !$, we obtain
\begin{eqnarray}  \label{sigma}
\frac{\Sigma \left( \xi \right) }{2\rho _{c}}= \frac{R}{\pi}%
\sum_{s=0}^{\infty }\sum_{l=0}^{2s+1}\frac{\left( -1\right) ^{s}2^{-\left(
2s+1\right) }C_{l}^{2s+1}}{\left( 2s+1\right) !\left( 2s-2l+1\right) }%
k^{2s+1}\xi ^{2s+1}  \notag \\
\times\left[ \left( \frac{R}{\xi }\right) ^{2s-2l+1}\left( 1+\sqrt{1-\frac{%
\xi ^{2}}{R^{2}}}\right) ^{2s-2l+1}-1\right] .
\end{eqnarray}
In the following, for the sake of clarity and simplicity, instead of showing
the general results that can be derived from Eq.~(\ref{sigma}), we will
present only the results obtained for some particular values of $s$. More
exactly, we will restrict our study of lensing to the specific case $s=7$.
The expansion of the $\sin x$ function for $s=7$ gives a very good
approximation of the function for $x$ in the range $0\leq x\leq\pi$. The
series expansions can be easily generalized and obtained for larger values
of $s$, thus giving the possibility of obtaining the lensing parameters of
Bose-Einstein condensate dark matter halos at any prescribed level of
precision. In fact, since the condition $x\leq \pi$ can be formulated as $%
\pi \xi\cosh \vartheta\leq \pi$, it follows that the adopted approximation
gives a very good description of the condensate dark matter halos for all $%
r\leq R$.

For $s=7$ the surface mass density of the dark matter halo is given by
\begin{eqnarray}  \label{sigma2}
\frac{\Sigma \left( \xi \right) }{2\rho _{c}}= \sqrt{R^{2}-\xi ^{2}}\Bigg[%
\left( 1-\frac{\pi ^{2}}{18}+\frac{\pi ^{4}}{600}-\frac{\pi ^{6}}{35\,280}%
\right)  \notag \\
-\frac{\pi ^{2}}{9}\left( 1-\frac{\pi ^{2}}{50}+\frac{3\pi ^{4}}{9800}%
\right) \left( \frac{\xi }{R}\right) ^{2}  \notag \\
+\frac{\pi ^{4}}{225}\left( 1-\frac{\pi ^{2}}{98}\right) \left( \frac{\xi }{R%
}\right) ^{4}-\frac{\pi ^{6}}{11025}\left( \frac{\xi }{R}\right) ^{6}\Bigg],
\end{eqnarray}
or, in numerical form,
\begin{eqnarray}
\Sigma \left( \xi \right) = \sqrt{R^{2}-\xi ^{2}}\Bigg[ 1.17357-1.82572%
\left( \frac{\xi }{R}\right) ^{2}  \notag \\
+0.778658\left( \frac{\xi }{R}\right) ^{4}-0.174402\left( \frac{\xi }{R}%
\right) ^{6}\Bigg] \rho _{c}.
\end{eqnarray}
In this approximation the central surface mass density can be evaluated as
\begin{equation}
\frac{\Sigma (0)}{2\rho _c}\approx R\left(1 -\frac{\pi ^2}{18}+\frac{\pi ^4}{%
600}-\frac{\pi ^6}{35280}\right)=0.5867R.
\end{equation}

The total mass of the Bose-Einstein condensate model can be calculated by
integrating the surface mass density over the plane of the sky, $M_{\xi
}\left(R\right)=2\pi \int_0^R{\Sigma (\xi) \xi d\xi }$, and is given by
\begin{eqnarray}
M_{\xi }(R)& \approx & \frac{4\pi }{3}\left( 1-\frac{\pi ^{2}}{10}+\frac{\pi
^{4}}{280}-\frac{\pi ^{6}}{15120}\right) \rho _{c}R^{3}  \notag \\
&=& 1.2455\rho _{c}R^{3},
\end{eqnarray}
which gives an excellent approximation for the exact mass formula $%
M(R)=(4/\pi)\rho _cR^3=1.2739\rho _cR^3$.

An important quantity for gravitational lensing studies is the cumulative
surface mass density, i.e., the total mass contained in a infinite cylinder
with radius $\xi $, $M\left(\xi \right)=2\pi \int_0^{\xi}{\Sigma \left(\xi
^{\prime }\right)\xi ^{\prime }d\xi ^{\prime }}$ \cite{Sch,Wamb, Ret}, which
for the Bose-Einstein condensate dark halo is given by
\begin{eqnarray}  \label{mxi}
M_{\xi }\left( \xi \right) =\rho _{c}\pi \xi ^{2}\sqrt{R^{2}-\xi ^{2}}\Bigg[%
\left( 2-\frac{\pi ^{2}}{9}+\frac{\pi ^{4}}{300}-\frac{\pi ^{6}}{17\,640}%
\right)  \notag \\
-\pi ^{2}\left( \frac{2}{9}- \frac{\pi ^{2}}{225}+\frac{\pi ^{4}}{14700}%
\right) \left( \frac{\xi }{R}\right) ^{2}  \notag \\
+\pi ^{4}\left( \frac{2}{225}-\frac{\pi ^{2}}{11025}\right) \left( \frac{\xi
}{R}\right) ^{4}-\frac{2\pi ^{6}}{11\,025}\left( \frac{\xi }{R}\right) ^{6}%
\Bigg],
\end{eqnarray}
or, equivalently,
\begin{eqnarray}
M_{\xi }\left( \xi \right) &=& 3.68689\xi ^{2}\sqrt{R^{2}-\xi ^{2}}\Bigg[ 1-
1. 5557\left( \frac{\xi }{R}\right) ^{2}  \notag \\
&& +0.66349\left( \frac{\xi }{R}\right) ^{4}- 0.14861\left( \frac{\xi }{R}%
\right) ^{6}\Bigg] \rho _{c}.
\end{eqnarray}


\subsection{The lens equation and the deflection of light}


In the following, we consider a simplified gravitational lens scenario
involving a point source and a circular symmetrical lens. The three basic
`plane' in thin lens approximation are the source $S$, the lens $L$, and the
observer $O$. Light rays emitted from the source are detected by the lens.
For a point-like lens, there will always be (at least) two images $S_1$ and $%
S_2$ of the source. With external shear (due to the tidal field of objects
outside but near the light bundles) there can be more images. The observer
sees the images in directions corresponding to the tangents to the real
incoming light paths \cite{Sch, Wamb,Ret}.

In the thin lens approximation, the lens equation for axially symmetric lens
is \citep{Sch, Wamb,Ret}
\begin{equation}
\eta= \frac{D_S}{D_L}\xi -D_{LS}\tilde {\alpha},
\end{equation}
where the quantities $\eta $ and $\xi $ are the physical positions of the
source in the source plane and an image in the image plane, respectively, $%
\tilde {\alpha}$ is the deflection angle, and $D_L$, $D_S$ and $D_{LS}$ are
the angular distances from observer to lens, from observer to source, and
from lens to source, respectively. By introducing the dimensionless position
$\beta =\eta /D_S$, $\theta=\xi/D_L$ and the dimensionless angle $\alpha
=\left(D_{LS}/D_S\right)\tilde {\alpha}$, the thin lens equation can be
written as
\begin{equation}
\beta =\theta - \alpha (\theta).
\end{equation}

In the circular-symmetric case the deflection angle is given as
\begin{eqnarray}
D_L\alpha \left( \xi \right) &=& \frac{2}{\xi }\int_{0}^{\xi }\frac{\Sigma
\left( \xi ^{\prime }\right) }{\Sigma _{crit}}d\xi ^{\prime }=\frac{2}{\xi }%
\int_{0}^{\xi }\xi ^{\prime }\kappa \left( \xi ^{\prime }\right) d\xi
^{\prime }  \notag \\
&=& \frac{1}{\pi \Sigma _{crit}}\frac{M_{\xi }\left( \xi \right) }{\xi },
\end{eqnarray}
where the convergence $\kappa $ is defined as $\kappa \left( \xi \right)
=\Sigma \left( \xi \right) /\Sigma _{crit}$, where $\Sigma _{crit}$ is the
critical surface density defined as $\Sigma _{crit}=c^{2}D_{S}/4\pi
GD_{L}D_{LS}$. The central convergence, $\kappa _{c}=\kappa \left( 0\right) $
\cite{Sch, Wamb, Ret}, a parameter that determines the lensing properties of
the Bose-Einstein condensate dark matter halo profiles, is given by the
relation
\begin{equation}
\kappa_{c} = {\frac{1.17357 \rho_c R }{\Sigma_{\mathrm{crit}} }} =\frac{%
1.17357 \pi M(R)}{ 4 R^2 \Sigma_{\mathrm{crit}}}.
\end{equation}
The dimensionless surface mass density can be given by
\begin{eqnarray}
\kappa(D_L\theta)&=& \kappa_c {\frac{\sqrt{R^2-D_l^2\theta^2}}{1.17357 R}} \Bigg[%
1.17357 - 1.82572 \left(\frac{D_L\theta}{R}\right)^2  \notag \\
&&+ 0.778658 \left({\frac{D_L\theta}{R}}\right)^4 -0.174402 \left({\frac{D_L\theta%
}{R}}\right)^6\Bigg].\nonumber\\
\end{eqnarray}

For a spherically symmetric lens that is capable of forming multiple images
of the source, one sufficient condition is $\kappa _c > 1$. In the case $%
\kappa _c<1$, only one image of the source is formed. Similarly to the case
of other non-singular profiles, such as the Einasto profile \cite{Ret}, the
Bose-Einstein condensate dark matter density profiles are not capable of
forming multiple images for any mass. Instead, the condition $\kappa _c > 1$
sets a minimum value for lens mass required to form multiple images.

The deflection potential $\psi \left( \xi \right) $ for a spherically
symmetric lens is given by \cite{Sch, Wamb, Ret}
\begin{equation}
\psi \left( \xi \right) =2\int_{0}^{\xi }\xi ^{\prime }\kappa \left( \xi
^{\prime }\right) \ln \left( \frac{\xi }{\xi ^{\prime }}\right) d\xi
^{\prime },
\end{equation}
and can be computed to give
\begin{eqnarray}
\psi \left( \xi \right) &=&\rho _{c}\Bigg\{\left( \frac{16}{9}-\frac{4\pi
^{2}}{25}+\frac{4\pi ^{4}}{735}-\frac{\pi ^{6}}{10206}\right) R^{3}  \notag
\\
&& +\sqrt{R^{2}-\xi ^{2}} \Bigg[-\left( \frac{16}{9}-\frac{4\pi ^{2}}{25}+%
\frac{4\pi ^{4}}{735}-\frac{\pi ^{6}}{10206}\right) R^{2}  \notag \\
&&+\left( \frac{4}{9}-\frac{2\pi ^{2}}{25}+\frac{\pi ^{4}}{7350}-\frac{\pi
^{6}}{714420}\right) \xi ^{2}  \notag \\
&&-\frac{\pi ^{2}}{225}\xi ^{2}\left( 4-\frac{2\pi ^{2}}{49}+\frac{\pi ^{4}}{%
2646}\right) \left( \frac{\xi }{R}\right) ^{2}  \notag \\
&& \hspace{-1.00cm} +\frac{4\pi ^{4}}{11025}\xi ^{2}\left( 1-\frac{\pi ^{2}}{%
162}\right) \left( \frac{\xi }{R}\right) ^{4}-\frac{4\pi ^{6}}{893025}\xi
^{2}\left( \frac{\xi }{R}\right) ^{6}\Bigg]  \notag \\
&& \hspace{-1.25cm} -\frac{1}{3}\left( 4-\frac{2\pi ^{2}}{5}+\frac{\pi ^{4}}{%
70}-\frac{\pi ^{6}}{3780}\right) R^{3}\ln \frac{2R}{R+\sqrt{R^{2}-\xi ^{2}}}%
\Bigg\}.
\end{eqnarray}
In a numerical form we have
\begin{eqnarray}
\psi \left( \xi \right) &=& \rho _{c}\Bigg\{\sqrt{R^{2}-\xi ^{2}}R^{2}\Bigg[ %
-0.63456+0.368622\left( \frac{\xi }{R}\right) ^{2}  \notag \\
&& -0.159404\left( \frac{\xi }{R}\right) ^{4}+ 0.0331881\left( \frac{\xi }{R}%
\right) ^{6}  \notag \\
&& -0.00430621\left( \frac{\xi }{R}\right) ^{8}\Bigg] + 0.63456R^{3}\times
\notag \\
&& \times \left( 1- 0.62478\ln \frac{2.R}{R+\sqrt{R^{2}-\xi ^{2}}}\right) %
\Bigg\}.  \label{eq:lenspsi}
\end{eqnarray}


\subsection{The magnification factor}


Gravitational lensing effect preserves the surface brightness but it causes
variations in shape and solid angle of the source. Therefore, the source
luminosity is amplified by a magnification factor $\mu $, given by \cite%
{Sch, Wamb, Ret}
\begin{equation}
\mu =\frac{1}{\left(1-\kappa \right)^2-\gamma ^2},
\end{equation}
where $\gamma =\gamma \left(\xi \right)$ is the shear. For a spherically
symmetric lens the shear is given by
\begin{equation}
\gamma \left(\xi \right)=\bar{\kappa }-\kappa=\frac{\bar{\Sigma }%
\left(\xi\right)-\Sigma \left(\xi \right)}{\Sigma _{crit}},
\end{equation}
where $\bar{\kappa }=\bar{\Sigma }\left(\xi \right)/\Sigma _{crit}$, and $%
\bar{\Sigma }\left(\xi \right)$ is the average surface mass density within $%
\xi $ given by
\begin{equation}
\bar{\Sigma}\left(\xi \right)=\frac{2}{\xi ^2}\int_0^{\xi }{\xi ^{\prime
}\Sigma \left(\xi ^{\prime }\right)d\xi ^{\prime }}.
\end{equation}

By using Eq.~(\ref{sigma2}) we obtain for $\bar{\Sigma}\left( \xi \right) $
the expression
\begin{eqnarray}
\bar{\Sigma}\left( \xi \right) &=&2R\rho _{c}\left( \frac{\xi }{R}\right)
^{-2}\Bigg\{0.198228+\sqrt{1-\left( \frac{\xi }{R}\right) ^{2}}\times  \notag
\\
&&\Bigg[0.487671\left( \frac{\xi }{R}\right) ^{2}-0.384069\left( \frac{\xi }{%
R}\right) ^{4}+  \notag \\
&&0.114005\left( \frac{\xi }{R}\right) ^{6}-\frac{0.019378\xi ^{8}}{R^{8}}%
-0.198228\Bigg]\Bigg\}.  \notag \\
\end{eqnarray}

Therefore for the shear $\gamma (\xi )$ of a BEC dark matter halo we obtain
the expression
\begin{eqnarray}
\gamma (\xi )&=&\frac{R\rho _{c}}{\Sigma _{crit}}\sqrt{1-\left( \frac{\xi }{R%
}\right) ^{2}}\Bigg[ -0.68590-\nonumber\\
&&0.198228\left( \frac{\xi }{R}\right) ^{-2}+
0.155028\left( \frac{\xi }{R}\right) ^{6}-\nonumber\\
&&0.664\,65\left( \frac{\xi }{R}%
\right) ^{4}+  
1.4417\left( \frac{\xi }{R}\right) ^{2}+\nonumber\\
&&\frac{0.198228}{\left( \xi
/R\right) ^{2}\sqrt{1-\left( \xi /R\right) ^{2}}} \Bigg ].
\end{eqnarray}

The magnification by a spherically symmetric lens can be written as \cite%
{Sch, Wamb, Ret}
\begin{eqnarray}  \label{mu}
\mu &=&\frac{1}{\left( 1-\bar{\kappa}\right) \left( 1+\bar{\kappa}-2\kappa
\right) }=  \notag \\
&&\frac{1}{\left( 1-\bar{\Sigma}(\xi)/\Sigma_{crit}\right) \left( 1+\bar{%
\Sigma}(\xi)/\Sigma _{crit}-2\Sigma (\xi)/\Sigma _{crit} \right) }.  \notag
\\
\end{eqnarray}

The magnification may be divergent for some image positions. The loci of the
diverging magnification in the image plane are called the critical curves.
From Eq.~(\ref{mu}) we see that the lensing profile has two critical curves.
The first curve, $1-\bar{\kappa }=0$ is the tangential critical curve, which
corresponds to an Einstein ring with a given Einstein radius. The second
curve, $1+\bar{\kappa}-2\kappa =0$ is the radial critical curve, which also
defines a ring, and its corresponding radius \citep{Sch, Wamb, Ret}.


\section{The virial theorem and the perturbation equation for Bose-Einstein
condensates}

\label{sect5}

One of the basic relations in theoretical physics, the virial theorem,
represents a very powerful method for the study of the equilibrium and
perturbation properties of fluids, including the quantum ones. The virial
theorem was originally proved, and used for the study of the equilibrium and
of the stability of rotating fluid bodies, bound by self-gravitation \cite%
{Chand1, Chand2, Chand3}. One of the important forms of the virial theorem,
the tensor-virial theorem, reduces the local hydrodynamical Euler equations
into global virial equations, which give the essential information on the
structure and stability of the whole gravitating system. An important
application of the virial theorem is for the study of the small
perturbations of incompressible uniform ellipsoids when perturbed from an
initial equilibrium state. In this case, in the absence of viscous
dissipation, each perturbed virial equation provides a different set of
normal modes \cite{Sed}. Moreover, in the physically very interesting case
when the equilibrium of the considered system is sustained by an external
confining potential, the perturbations of the fluid obviously do not modify
the confining potential itself. Therefore it follows that the tensor virial
methods, and the virial theorems, can be extended to nonuniform compressible
flows. The virial theorem represent an extremely powerful method for the
study of gases with polytropic equations of state \cite{Chand1, Chand2,
Chand3,Tas1, Chand4,Tas2}. It has also been extensively used in the study of
Bose-Einstein Condensates \citep{Lush, Pit, Sed}.

\subsection{The scalar virial theorem}

In order to derive the scalar virial theorem we consider the behavior of the
physical parameters of the dark matter halos under the scaling
transformation $\vec{r}\rightarrow \alpha \vec{r}$, where $\alpha $ is a
constant. Then the normalization condition giving the total number of
particles, $N=\int \left| \psi \left( \vec{r}\right) \right| ^{2}d^3\vec{r}$
requires that $\psi \left(\vec{r}\right)\rightarrow \alpha ^{-3/2} \psi
\left(\vec{r}\right)$ \cite{Da99,rev6}. Thus the total energy scales as
\begin{equation}
E\left[\alpha \right]=\alpha ^{-2}E_K+\alpha ^2E_{rot}+\alpha
^{-3}E_{int}+\alpha ^{-1}E_{grav}.
\end{equation}
Since the energy is stationary for any variation of the wave function $\psi $
around the exact solution of the Gross-Pitaevskii equation, by requiring the
energy variation to vanish at first order in $\alpha $, that is, $%
\left.\left( \delta E\left[ \alpha \right] /\delta \alpha \right) \right|
_{\alpha =1}=0$, we obtain the virial theorem in the form \cite{Da99, rev6}
\begin{equation}
2E_K-2E_{rot}+3E_{int}+E_{grav}=0.
\end{equation}

By multiplying both sides of Eq.~(\ref{poin}) with $\rho \left(\vec{r}\right)$
and integrating over $\vec{r}$ we obtain
\begin{equation}  \label{75}
\mu N=E_{rot}+2E_{int}+2E_{grav},
\end{equation}
where the interaction energy $E_{int}$ and the gravitational energy $E_{grav}$ are defined as $E_{int}=\left(U_0/2\right)\int_V{\rho ^2(\vec{r})d^3\vec{r}}$, and $V_{grav}=\left(m_{\chi}^2/2\right)\int_V{V_{grav}\left(\vec{r}\right)\rho \left(\vec{r}\right)d^3\vec{r}}$, respectively.
On the other hand from Eq.~(\ref{poin}) we obtain
\begin{equation}  \label{76}
\mu =\frac{U_0}{m_{\chi}}\rho _c+m_{\chi}V_{grav}(0).
\end{equation}
By using Eqs.~(\ref{75}) and (\ref{76}) we obtain for the total energy of
the dark matter halo the expression
\begin{equation}
E=\frac{1}{2}\left[\frac{U_0}{m_{\chi }}\rho _c+m_{\chi }V_{grav}(0)\right]+%
\frac{1}{2}E_{rot}.
\end{equation}

By using the virial theorem we can easily find the condition for the
validity of the Thomas-Fermi approximation. By using the density for the
static condensate given by Eq.~(\ref{neq1}), it is easy to estimate $%
E_K\approx \bar 2M(R)k^2/m_{\chi}^2$ and $E_{int}\approx
U_0M(R)^2k^3/m_{\chi }^2$, respectively. The Thomas-Fermi approximation
requires $E_K<<E_{int}$, giving
\begin{equation}
N=\frac{M}{m_{\chi }}>>\frac{1}{kl_a}=\frac{R}{\pi l_a},R>>\sqrt{\frac{%
m_{\chi }}{4l_{a}\rho _{c}}}.
\end{equation}
Hence for systems with enough high particle numbers the Thomas-Fermi
approximation is always valid.

\subsection{The tensor virial theorem}

In the following we assume that the dark matter gravitational condensate
rotates like a rigid body, and its rotation is supported by an array of
vortices. Moreover, it is confined by a gravitational field with potential $%
V_{grav}$, assumed to satisfy the Poisson equation Eq.~(\ref{poi}). To
obtain the tensor virial equation for the BEC halo we multiply both sides of
Eq.~(\ref{6}) by $x_{i}$ and we integrates over the volume $V$ of the dark
matter halo. Thus we first obtain \cite{Sed, Chand1, Chand2, Chand3}
\begin{equation}
\frac{d}{dt}\int_{V}\rho x_{i}u_{j}d^{3}\vec{r}=\int_{V}\rho x_{i}\frac{%
du_{j}}{dt}d^{3}\vec{r}+\int_{V}\rho u_{i}u_{j}d^{3}\vec{r}.
\end{equation}
The last term on the right-hand of Eq.~(\ref{6}) can be rewritten as
\begin{equation}
\frac{1}{2}\int_{V}\rho x_{i}\frac{\partial }{\partial x_{j}}\left| \vec{%
\Omega}\times \vec{r}\right| ^{2}d^{3}\vec{r}=\vec{\Omega}^{2}I_{ij}-\Omega
_{j}I_{ik}\Omega _{k},
\end{equation}
where
\begin{equation}
I_{ij}=I_{ji}=\int_{V}\rho x_{i}x_{j}d^{3}\vec{r},
\end{equation}
is the moment of inertia tensor of the dark matter condensate. Assuming that
the pressure becomes a constant $p_{0}$ outside the dark matter
distribution, one can write
\begin{eqnarray}
-\int_{V}x_{i}\frac{\partial p}{\partial x_{j}}d^{3}\vec{r}%
&=&-\int_{S}x_{i}p_{0}d\Sigma _{j}+\int_{V}\delta _{ij}pd^{3}\vec{r}=  \notag
\\
&&\delta _{ij}\int_{V}\left( p-p_{0}\right) d^{3}\vec{r}= \delta _{ij}\bar{%
\Pi},
\end{eqnarray}
where $\delta _{ij}$ is the Kronecker delta symbol. By taking into account
the explicit form of the condensate self-gravitational potential, $\phi
\left( \vec{r},t\right) =-G\int_{V}d^{3}\vec{r}^{\prime }\rho \left( \vec{r}%
^{\prime },t\right) /\left| \vec{r}-\vec{r}^{\prime }\right| $, one can show
that
\begin{equation}
\int_{V}\rho x_{i}\frac{\partial \phi }{\partial x_{j}}d^{3}\vec{r}=-\frac{1%
}{2}\int_{V}\rho B_{ij}d^{3}\vec{r}=-\Phi _{ij},  \label{8}
\end{equation}
where
\begin{equation}
B_{ij}\left( \vec{r},t\right) =-G\int_{V}\rho \left( \vec{r}^{\prime
},t\right) \frac{\left( x_{i}-x_{i}^{\prime }\right) \left(
x_{j}-x_{j}^{\prime }\right) }{\left| \vec{r}-\vec{r}^{\prime }\right| ^{3}}%
d^{3}\vec{r}^{\prime },
\end{equation}
is the potential tensor of Chandrasekhar \cite{Chand1, Chand2,Chand3}. By
defining the kinetic energy tensor as
\begin{equation}
T_{ij}= \frac{1}{2} \int_{V}\rho u_{i}u_{j}d^{3}\vec{r},
\end{equation}
the tensor virial theorem for a rotating Bose-Einstein Condensate dark
matter halo in a gravitational field can be written as \cite{Chand1,
Chand2,Chand3, Tas1}
\begin{eqnarray}  \label{vir9}
\frac{d}{dt}\int_{V}\rho x_{i}u_{j}d^{3}\vec{r}&=&2T_{ij}+\delta _{ij}\bar{%
\Pi}+\Phi _{ij}+\vec{\Omega}^{2}I_{ij}-  \notag \\
&&\Omega _{j}I_{ik}\Omega _{k}+2\varepsilon _{jkl}\Omega _{k}\int_{V}\rho
x_{i}u_{l}d^{3}\vec{r}.  \notag \\
\end{eqnarray}

For a dark matter halo in hydrostatic equilibrium $u_{i}=0$, and from Eq.~(%
\ref{vir9}) we find
\begin{equation}
\Phi _{ij}+\delta _{ij}\bar{\Pi}+\vec{\Omega}^{2}I_{ij}-\Omega
_{j}I_{ik}\Omega _{k}=0.
\end{equation}

For $i\neq j$ we have $\Phi _{ij}+\vec{\Omega}^{2}I_{ij}-\Omega
_{j}I_{ik}\Omega _{k}$, while contracting over $i$ and $j$ gives
\begin{equation}
\Phi +3\bar{\Pi}+\vec{\Omega}^{2}I-\Omega _{i}I_{ij}\Omega _{j}=0,
\label{virrot}
\end{equation}
where $\Phi $ represents the gravitational potential energy.

\subsubsection{Gravitational energy of a rotating dark matter halo}

As an application of the tensor virial theorem we determine in the following
the gravitational energy of a rotating dark matter halo. If the $z$-axis is
chosen as axis of rotation, Eq. (\ref{virrot}) takes the form
\begin{equation}
\Phi +3\bar{\Pi}+\vec{\Omega}^{2}I_{\Omega }=0,  \label{virec}
\end{equation}
where $I_{\Omega }=\int_{V}\rho \left( x^{2}+y^{2}\right) d^{3}\vec{r}$. For
the case of the rotating Bose-Einstein condensate dark matter halo the
equation of the hydrostatic equilibrium takes the form
\begin{equation}
2\nabla \frac{p}{\rho }=\nabla \left( V_{grav}+\frac{1}{2}\left| \vec{\Omega}%
\times \vec{r}\right| ^{2}\right) ,
\end{equation}
and can be integrated to give
\begin{equation}
2p=\rho \left( V_{grav}+\frac{1}{2}\left| \vec{\Omega}\times \vec{r}\right|
^{2}-V_{grav}^{(0)}\right) ,  \label{p}
\end{equation}
where $V_{grav}^{(0)}$ is the gravitational potential at the pole of the
dark matter halo. Integrating Eq. (\ref{p}) over the dark matter halo volume
we obtain
\begin{equation}
2\bar{\Pi}=-2\Phi +\frac{1}{2}\vec{\Omega}^{2}I_{\Omega }-V_{grav}^{(0)}M,
\end{equation}
where $\Phi =-\left( 1/2\right) \int \rho V_{grav}d^{3}\vec{r}$. Hence from
Eq.~(\ref{virec}) we obtain \cite{Chand4}
\begin{equation}  \label{94}
\Phi =\frac{1}{4}\left( \frac{7}{2}\vec{\Omega}^{2}I_{\Omega
}-3V_{grav}^{(0)}M\right) .
\end{equation}

For a static halo $\vec{\Omega}=0$, $V_{grav}^{(0)}=GM/R$, and we reobtain
the well-known result $\Phi =-(3/4)GM^{2}/R$.

\subsection{The perturbation equation}

We consider now the small oscillations about equilibrium of a Bose Einstein
Condensate dark matter halo. We assume that in the stationary state there
are no fluid motions. In order to obtain the perturbation equation we use
the tensor virial theorem. Considering only periodic oscillations with
frequency $\omega $, representing the most interesting case from a physical
point of view, the Lagrangian displacement of a mass element $dm$ can be
written as $\vec{\xi}\left( \vec{r},t\right) =\vec{\xi}\left( \vec{r}\right)
\exp \left( i\sigma t\right) $ \cite{Sed, Chand1,Chand2,Chand3, Tas1,Chand4}%
. Due to the equation of continuity for such Lagrangian displacements $dm$
is a constant, and $\delta \rho /\rho +\nabla \cdot \vec{\xi}=0$. By taking
the Eulerian variation of the tensor virial equation for the condensate dark
matter halo we obtain
\begin{eqnarray}
\delta \frac{d}{dt}\int_{V}\rho x_{i}u_{j}d^{3}\vec{r}&=&2\delta
T_{ij}+\delta _{ij}\delta \bar{\Pi}+\delta \Phi _{ij}+\vec{\Omega}^{2}V_{ij}
\notag \\
  && \hspace{-1.5cm} -\Omega _{j}\Omega _{k}V_{kj}+2\varepsilon _{jkl}\Omega _{k}\delta
\int_{V}\rho x_{i}u_{l}d^{3}\vec{r},  
\end{eqnarray}
where
\begin{equation}
V_{ij}=\delta I_{ij}=\int_{V}\rho \left( \xi _{i}x_{j}+\xi _{j}x_{i}\right)
d^{3}\vec{r},
\end{equation}
is a tensor symmetric in its indices. By defining its non-symmetric part as $%
V_{i;j}=\int_{V}\rho \xi _{i}x_{j}d^{3}\vec{r}$, we have $%
V_{ij}=V_{i;j}+V_{j;i}$. For the variation of $\Phi _{ij}$ we find the
equation
\begin{eqnarray}
\delta \Phi _{ij} &=&-G\int_{V}\int_{V}dmdm^{\prime }\xi _{k}\frac{\partial
}{\partial x_{k}}\frac{\left( x_{i}-x_{i}^{\prime }\right) \left(
x_{j}-x_{j}^{\prime }\right) }{\left| \vec{r}-\vec{r}^{\prime }\right| ^{3}}
\notag \\
&=&-G\int_{V}d^{3}\vec{r}\rho \left( \vec{r}\right) \xi _{k}\frac{\partial }{%
\partial x_{k}}\int_{V}d^{3}\vec{r}^{\prime }\rho \left( \vec{r}^{\prime
}\right) \times  \notag \\
&& \hspace{-1.2cm} \xi _{k}\frac{\partial }{\partial x_{k}}\frac{\left( x_{i}-x_{i}^{\prime
}\right) \left( x_{j}-x_{j}^{\prime }\right) }{\left| \vec{r}-\vec{r}%
^{\prime }\right| ^{3}} -\int_{V}d^{3}\vec{r}\rho \left( \vec{r}\right) \xi
_{k}\frac{\partial B_{ij}}{\partial x_{k}}, 
\end{eqnarray}
which gives the important first order change in the gravitational potential
energy, due to a small perturbation of the matter in the condensate dark
matter system. Therefore, when there are no condensate dark matter motions
in the unperturbed frame, the second order virial equation takes the form
\cite{Sed, Chand1, Chand2,Chand3}
\begin{equation}
\frac{d^{2}V_{i;j}}{dt^{2}}=2\varepsilon _{ikl}\Omega _{l}\frac{dV_{k;j}}{dt}%
+\delta \Phi _{ij}-\Omega _{i}\Omega _{k}V_{kj}+\delta _{ij}\delta \bar{\Pi}.
\end{equation}

For small periodic perturbations we obtain \cite{Sed}
\begin{equation}
\sigma ^{2}V_{i;j}=2\varepsilon _{ikl}\Omega _{l}\sigma V_{k;j}+\delta \Phi
_{ij}-\Omega _{i}\Omega _{k}V_{kj}+\delta _{ij}\delta \bar{\Pi}.
\label{fpert}
\end{equation}

Eq.~(\ref{fpert}) contains all the second-harmonic modes of the rotating
condensate dark matter halo confined by its own gravitational field \cite%
{Sed, Chand1, Chand2,Chand3}.

\section{Applications of the virial theorem to BEC dark matter halos}
\label{6n}

As a simple example of the application of the virial theorem for the study
of the stability of the Bose-Einstein Condensate dark matter structures we
consider the case of the slowly rotating and slightly distorted galactic
halos. By employing the Lagrangian variables and the scalar virial theorem,
and by using the linear approximation, we obtain the angular frequency $%
\sigma $ of the lowest radial mode, as well as the condition for dynamical
instability, depending on the numerical value of the adiabatic exponent $%
\gamma =1+1/n$, where $n$ is the polytropic index.

\subsection{The scalar virial theorem for axisymmetric BEC dark matter halos}

We consider the BEC dark matter halo as a uniformly rotating, homogeneous
and compressible spheroid. The galactic halo rotates at $t=0$ with an
angular velocity $\Omega \left(t_0\right)=\Omega _0$. We restrict our
analysis to axisymmetric pulsations, and work in a Cartesian coordinate
system $\left(x_1,x_2,x_3\right)$, which rotates at any instant $t$ with the
instantaneous angular velocity $\Omega (t)$ of the galactic halo. Moreover,
we assume that $\Omega $ is oriented along the $x_3$ axis. We will also
ignore dissipative and electromagnetic effects. Under such assumptions, the
equation of motion of the galactic halo, Eq.~(\ref{6}), can be written as
\cite{Chand1, Chand2, Chand3, Tas1,Chand4, Tas2}
\begin{eqnarray}  \label{mot1}
\frac{du_i}{dt}+2\epsilon _{ik3}\Omega u_k+\epsilon_{ik3}\frac{d\Omega }{dt}%
x_k&=&-\frac{1}{\rho }\frac{\partial p}{\partial x_i}-\frac{\partial V_{grav}%
}{\partial x_i}  \notag \\
&& \hspace{-1cm} +\left(1-\delta _{i3}\right)\Omega ^2x_i,  
\end{eqnarray}
where $i=1,2,3$ and we have taken into account that the angular velocity $\Omega $
explicitly depends on time. For axisymmetric motions all terms in Eq.~(\ref%
{mot1}) describe motions in meridional planes, except the last two terms on
the left-hand side of the equation \cite{Chand4,Tas1}. Therefore Eq.~(\ref%
{mot1}) can be split as
\begin{equation}  \label{mot2}
\frac{du_i}{dt}=-\frac{1}{\rho }\frac{\partial p}{\partial x_i}-\frac{%
\partial V_{grav}}{\partial x_i}+\left(1-\delta _{i3}\right)\Omega ^2x_i,
\end{equation}
where $i=1,2,3$ and
\begin{equation}  \label{mot3}
\epsilon _{ik3}\left(2\Omega u_k+\frac{d\Omega }{dt}x_k\right)=0,
\end{equation}
respectively. Equation~(\ref{mot3}) can be immediately integrated to give
\begin{equation}  \label{mot4}
\frac{d}{dt}\left[\Omega \left(x_1^2+x_2^2\right)\right]=0.
\end{equation}

Equation~(\ref{mot4}) is the equation of conservation of the angular momentum for
each fluid element, a property which is true for axisymmetric motions only.
Now we multiply Eq.~(\ref{mot2}) by $x_i$, we integrate over the total mass $%
M$ of the galactic halo, and add the resulting equations. Hence we obtain
the scalar virial theorem for the BEC halo in the form \cite{Chand4,Tas1}
\begin{eqnarray}  \label{vir1}
\frac{1}{2}\frac{d^2}{dt^2}\int_M{x_kx_kdm}=\int_M{\frac{dx_k}{dt}\frac{%
dx_k}{dt}dm}+\Phi   \notag \\
+3\int_M{\frac{p}{\rho}dm}+\int_M{\Omega\left(x_1^2+x_2^2\right)dm},
\end{eqnarray}
where
\begin{equation}  \label{potn}
\Phi=-\frac{1}{2}G\int_M\int_M{\frac{dmdm^{\prime }}{|\vec{r}-\vec{r}%
\;^{\prime }|}},
\end{equation}
is the gravitational potential energy of the dark matter halo. At time $t=0$
the dark matter halo is in relative equilibrium, and the virial theorem Eq.~(%
\ref{vir1}) gives
\begin{equation}  \label{eqcond}
\Phi _0+3\int_M{\frac{p_0}{\rho_0}dm}+\int_M{\Omega_0\left(a_1^2+a_2^2%
\right)dm}=0,
\end{equation}
where the subscript zero indicates the initial values, and $a_1$ and $a_2$
represents the $t=0$ values of the coordinates $x_1$ and $x_2$, respectively.

\subsection{The equations of motion in Lagrangian coordinates}

It is more convenient to study the axisymmetric motion of the galactic dark
matter halos in Lagrangian coordinates, in which all variables
characterizing the BEC system, $x_i$, $\rho $, $p$, and $V$ are expressed as
functions of the independent quantities $a_i$ and $t$ \cite{Chand4,Tas2}.
Therefore the Eulerian coordinates $x_i$ can be expressed as $%
x_i=x_i\left(a_i,t\right)$. Hence, when axisymmetry is preserved, there are
expanding and contracting solutions for which the coordinates of a particle
are functions of their initial values and of the time only.

In Lagrangian coordinates the equations of motion of the BEC dark matter
halo Eqs.~(\ref{mot2}) can be reformulated as \cite{Tas2}
\begin{equation}
\frac{\partial x_i}{\partial a_i}\frac{\partial ^2x_i}{\partial t^2}=-\frac{1%
}{\rho}\frac{\partial p}{\partial a_i}-\frac{\partial V_{grav}}{\partial a_i}%
+\left(1-\delta _{i3}\right)\Omega ^2x_i\frac{\partial x_i}{\partial a_i},
\end{equation}
where $i=1,2,3$.

The conservation of mass imposes the condition
\begin{equation}  \label{consn}
\rho J=\rho _0,
\end{equation}
where $J$ is the Jacobian of the coordinate transformation \cite{Tas2},
\begin{equation}
J=\frac{\partial x_1}{\partial a_1}\frac{\partial x_2}{\partial a_2}\frac{%
\partial x_3}{\partial a_3}.
\end{equation}
As for the pressure of dark matter halo, we will assume that it is given by
Eq.~(\ref{pressn}), that is, by a polytropic equation of state with
polytropic index $n=1$. We assume that the pressure vanishes on the moving
surface, at the vacuum boundary of the dark matter halo. Moreover, the
gravitational potential must be continuous across the galactic boundary.
Once these conditions are satisfied, axisymmetric motions are then entirely
determined once an initial velocity distribution is prescribed at every
point \cite{Chand4}.

\subsection{Pseudo-radial oscillations of BEC dark matter halos}

In the following we restrict our analysis to the case of pseudo-radial
pulsations of slowly rotating and slightly distorted galactic polytropic
dark matter halos. For slowly rotating spheroids in the first approximation
one can adopt for the evolution of the coordinates a linear expression, so
that \cite{Tas2}
\begin{equation}
x_i=a_i\zeta (t), \qquad i=1,2,3.
\end{equation}
The equation governing the time evolution of $w(t)$ can be obtained from the
scalar virial theorem Eq.~(\ref{vir1}) in Lagrange coordinates. First of
all, by using the adopted representations of $x_i$ we obtain immediately the
identity
\begin{equation}
\frac{1}{2}\frac{d^2}{dt^2}\int_M{x_kx_kdm}-\int_M{\frac{dx_k}{dt}\frac{dx_k%
}{dt}dm}=\zeta \frac{d^2\zeta }{dt^2}I_0,
\end{equation}
where $I_0$ denotes the initial moment of inertia of the rotating galaxy,
and we have used the constancy of the mass element $dm$ as we follow the
motion of the particles. In order to obtain the time variation of the
gravitational potential energy, given by Eq.~(\ref{potn}), we take into
account that $\left |\vec{r}-\vec{r}^{\prime }\right |=\zeta (t)\left |\vec{a%
}-\vec{a}^{\prime }\right |$, in which $\vec{a}=\vec{r}(0)$ and $\vec{a}%
^{\prime }=\vec{r}^{\prime }(0)$ \cite{Tas2}. Then we obtain for the
gravitational potential energy the expression
\begin{equation}
\Phi =\frac{\Phi _0}{\zeta ^2}.
\end{equation}
With the use of Eq.~(\ref{consn}) and of the equation of state of the BEC
dark matter we obtain
\begin{equation}
\frac{p}{\rho }=\frac{1}{\zeta ^3}\frac{p_0}{\rho _0}.
\end{equation}
By integrating the above equation, we obtain
\begin{equation}  \label{intc}
3\int_M{\frac{p}{\rho }dm}=\frac{1}{\zeta ^3}\left(\left |\Phi _0\right
|-2K_0\right),
\end{equation}
where $K_0$ represents the initial rotational kinetic energy, and in order
to eliminate the integral over pressure in Eq.~(\ref{intc}) we have used the
equilibrium condition given by Eq.~(\ref{eqcond}) \cite{Tas2}. The
conservation of the angular momentum immediately gives the equation
\begin{equation}
\Omega =\frac{\Omega _0}{\zeta ^2},
\end{equation}
and, consequently,
\begin{equation}
\int_M{\Omega ^2\left(x_1^2+x_2^2\right)dm}=\frac{2K_0}{\zeta ^2}.
\end{equation}

Therefore, with the use of the above results the scalar virial theorem Eq.~(%
\ref{vir1}) we obtain the time evolution of $\zeta $ as \cite{Tas2}
\begin{equation}  \label{117}
\frac{d^2\zeta }{dt^2}=\frac{1}{\zeta ^2}\left(\frac{1}{\zeta ^2}-1\right)%
\frac{\left |\Phi _0\right|}{I_0}+\frac{1}{\zeta ^3}\left(1-\frac{1}{\zeta }%
\right)\frac{2K_0}{I_0}.
\end{equation}
Eq.~(\ref{117}) must be solved with the appropriate boundary conditions. It
is also important to mention that the quantities $\Phi _0$, $K_0$ and $I_0$
refer to the rotating galactic halo. Moreover, in obtaining Eq.~(\ref{117})
we have assumed that the halo rotates as a solid body.

\subsection{Oscillation frequency of BEC dark matter halos in the linear
approximation}

In the following we consider the small amplitude pseudo-radial oscillations
of the $n=1$ polytropic dark matter halos. We can see from Eq.~(\ref{117})
that the point $\zeta =1$ defines an equilibrium state of the galaxy. In
order to to describe linear and quasi-linear oscillations about this
equilibrium point, we Taylor expand $\zeta $ near $\zeta =1$. In the first
approximation we obtain
\begin{equation}
\zeta (t)=1+\epsilon (t),
\end{equation}
where $\epsilon (t)\ll 1$. In this approximation Eq.~(\ref{117}) becomes \cite%
{Tas2}
\begin{equation}  \label{119}
\frac{d^2\epsilon}{dt^2}+\sigma ^2\epsilon =\lambda \epsilon ^2+O(\epsilon
^3),
\end{equation}
where \cite{Chand4,Tas2}
\begin{equation}  \label{122}
\sigma ^2=2\frac{\left |\Phi _0\right |}{I_0}-\frac{2K_0}{I_0},
\end{equation}
and
\begin{equation}
\lambda =\frac{1}{2}\left(7\sigma ^2-\frac{2K_0}{I_0}\right).
\end{equation}
If we neglect the second and third order powers of $\epsilon$, the solution
of Eq.~(\ref{119}) is obtained as
\begin{equation}
\epsilon (t)=\epsilon _0\sin \left(\sigma t\right).
\end{equation}
In terms of the Eulerian variables we obtain $\xi _i=\epsilon _0x_i\sin
(\omega t)$.

Equation (\ref{122}) gives the stability condition for BEC dark matter halos in
the linear approximation. If the halo is initially non-rotating, we have $%
K_0=0$, and the stability condition reduces to
\begin{equation}
\sigma ^2=2\frac{\left |\Phi _0\right |}{I_0}>0,
\end{equation}
a condition which is always satisfied by BEC dark matter halos described by
a polytropic equation of state with $n=1$ and $\gamma =2$, respectively.
With the use of Eq.~(\ref{94}) we can estimate the initial gravitational
energy of the dark matter halo as $\left |\Phi _0\right |=(3/4)GM^2/R$,
while for the moment of inertia we use the expression $I_0=2MR^2/5$ \cite%
{Chand4}. Then the oscillation frequency of the halo is given by
\begin{equation}
\sigma ^2=\frac{15GM}{8R^3}=\frac{15G}{2\pi}\rho _c,
\end{equation}
where we have taken into account Eq.~(\ref{26}) giving the mass-radius
relation of the static BEC dark matter halo. For the period of the
oscillations we obtain
\begin{eqnarray}
T=\frac{2\pi}{\sigma }&=&\sqrt{\frac{8}{15}}\pi ^{3/2}\frac{1}{\sqrt{G\rho _c%
}}=1.5745\times 10^{16}\times  \notag \\
&&\left(\frac{\rho _c}{10^{-24}\;\mathrm{g/cm^3}}\right)\;\mathrm{s}.
\end{eqnarray}

In the case of perturbed BEC dark matter halos with an initial rotation, the
stability condition reduces to
\begin{equation}
\left\vert \Phi _{0}\right\vert >K_{0},
\end{equation}%
or, equivalently, with the use of Eq.~(\ref{94}),
\begin{equation}
\frac{GM^{2}}{R}>2I_{\Omega _{0}}\Omega _{0}^{2},
\end{equation}%
where $\Omega _{0}$ is the initial angular velocity, and $I_{\Omega _{0}}$
the initial moment of inertia of the galactic halo. The period of
oscillations of a slowly rotating slightly disturbed BEC dark matter halo is
found as
\begin{eqnarray}
T&=&\frac{\sqrt{2}\pi \sqrt{I_{\Omega _{0}}}}{\sqrt{\left\vert \Phi
_{0}\right\vert -K_{0}}}\approx \frac{\sqrt{2}\pi \sqrt{I_{\Omega _{0}}}}{%
\sqrt{\left\vert \Phi _{0}\right\vert -K_{0}}}
   \nonumber \\
&=&\frac{\sqrt{32}\pi \sqrt{%
I_{\Omega _{0}}}}{\sqrt{2}\sqrt{2\Omega _{0}^{2}I_{\Omega _{0}}+GM^{2}/R}}.
\end{eqnarray}

By using again the mass-radius relation for BEC dark matter halos
given by Eq.~(\ref{26}), we obtain
\begin{equation}
T\approx \sqrt{\frac{8}{15}}\pi ^{3/2}\frac{1}{\sqrt{G\rho _{c}}}\frac{1}{%
\sqrt{1+15G\rho _{c}\Omega _{0}^{2}/\pi }}.
\end{equation}
This equation gives the corrections to the oscillations period of the halo
due to the presence of an initial rotation


\section{Discussions and final remarks}

\label{sect6}

The $n=1$ polytropic Bose-Einstein condensate dark matter model is the
simplest existing dark matter model. All the properties of the dark matter
distributions are determined by two parameters only: the mass and the radius
of the dark matter halo. The central density of the dark matter is
determined uniquely by $M(R)$ and $R$. For halos not contaminated with
baryonic matter both quantities must have the same universal value. However,
the presence of the baryonic matter may increase the size of the galactic
halo, thus leading to some variations in the total mass and radius of the
galactic structures.

Since all the properties of the Bose-Einstein condensates are determined by
two observational parameters only, once these parameters are known, the
physical properties of the galactic halos can be \textit{predicted} by the
model. Thus, for example, the tangential velocity of the test particles in
circular orbits around galaxies is determined by the \textit{universal}
equation
\begin{equation}
V^2(r)=\frac{GM}{R}\left[\frac{\sin (\pi r/R)}{\pi r/R}-\cos\frac{\pi R}{R}%
\right].
\end{equation}

\begin{figure}[tbp]
\centering
\includegraphics[width=8.15cm]{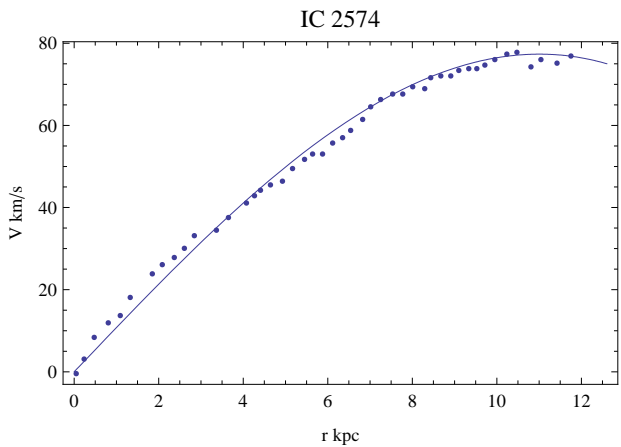}
\caption{Comparison of the predicted rotation curve of the dwarf galaxy IC
2574 (solid curve) with the observational data presented in \protect\cite{Oh}%
. The assumed mass of the galaxy is $M=1.64\times 10^{10}M_{\odot}$ (the
observationally determined value $M=1.462\times 10^{10}M_{\odot}$), and the
radius of the galaxy is $R=12.6$ kpc. }
\label{fig1}
\end{figure}

Once the mass of the galaxy and its radius is known, the rotation curves can
be obtained immediately. For example, in the case of the dwarf galaxy IC
2574, by assuming for the condensate component a mass of $M=1.64\times
10^{10}M_{\odot}$ and a radius of $R=12.6$ kpc, the rotation curve is given
by
\begin{equation}
V^2(r)=5.6374\times 10^3 \left[\frac{\sin (\pi r/R)}{\pi r/R}-\cos\frac{\pi R%
}{R}\right]\;\mathrm{km^2/s^2}.
\end{equation}
The comparison of the rotation curve \textit{predicted} by the Bose-Einstein
condensate dark matter model for the galaxy IC 2574 with the observational
data obtained in \cite{Oh} is represented in Fig.~\ref{fig1}.
The comparison of the \textit{predicted} rotation curve for the dwarf galaxy
M81 dwB with the observational data of \cite{Oh} is represented in Fig.~\ref%
{fig2}.
\begin{figure}[tbp]
\centering
\includegraphics[width=8.15cm]{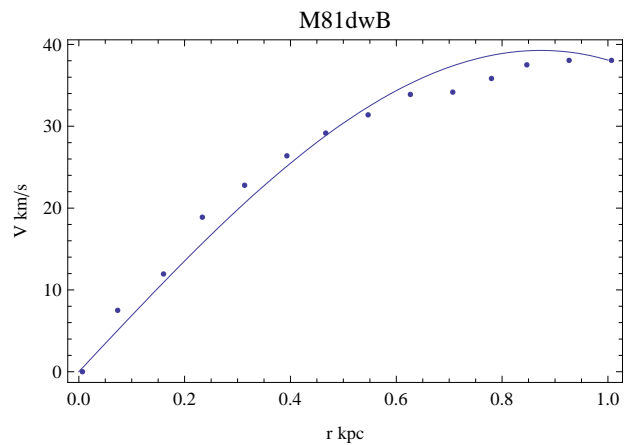}
\caption{Comparison of the rotation curve of the dwarf galaxy M81 dwB
predicted by the condensate dark matter model (solid curve) with the
observational data presented in \protect\cite{Oh}. The assumed mass of the
galaxy is $M=0.33\times 10^{9}M_{\odot}$ (the observationally determined
value $M=0.3\times 10^{9}M_{\odot}$), and the radius of the galaxy is $R=1$
kpc. }
\label{fig2}
\end{figure}

The condensate dark matter model \textit{predicts} a mean logarithmic
density slope of the inner core $\langle\alpha _{DM}\rangle=0.3068$, while
the observed value obtained from the study of brown galaxies is $%
\langle\alpha _{DM}\rangle=-0.29\pm0.07$ \citep{Oh} (the minus sign appears
due to the opposite sign convention adopted in the present paper). In \cite%
{Gent} it was shown, by using the mass decomposition of rotation curves
using cored haloes, that the product of the central density $\rho _c$ and
the core radius $R_{core}$ is a universal constant, independent of the
galaxy mass. In the case of condensate dark matter halos the product of the
central density and core can be written as
\begin{equation}
\rho _cR_{core}=0.00827 \left(\frac{M(R)}{10^{10}\;M_{\odot}}\right)\left(%
\frac{R}{10\;\mathrm{kpc}}\right)^{-2}\; \mathrm{g/cm^2},
\end{equation}
and for pure condensate dark matter halos this product must be a universal
constant. However, the condensate dark matter model predicts a dependence
\textit{on both mass and radius} of the constant. The presence of baryonic
matter may determine some modifications and variations in its numerical
value (for a comparison of the theoretical predictions with the
observational data see \cite{Mat}). The condensate dark matter model also
predicts a relation between $\rho _cR_{core}$ and the central convergence $%
\kappa _c$ of the halo of the form
\begin{equation}
\rho _cR_{core}=\frac{1}{7.372}\Sigma _{crit}R\kappa _c.
\end{equation}

Hence the condensate dark matter model is the only existing model that makes
easily testable \textit{predictions}, without any need for fitting, and
hence it can be easily falsified by comparing the theoretical predictions
with the observations. However, despite its remarkable successes at the
galactic scale, the Bose-Einstein condensate dark matter model needs further
testing. One of the best possibilities in confirming/ruling out the model is
through gravitational lensing. In the present paper we have obtained for the
first time all the relevant quantities necessary for an in depth comparison
of the theoretical predictions with the observational data in an analytical,
easy to handle form. These general analytical formulae enable arbitrary
precision calculation, as well as the study of the asymptotic behavior near
the vacuum boundary of the condensate. These formulae can be used in strong
and weak lensing studies of galaxies and clusters of galaxies, where dark
matter is assumed to be the dominant component. Moreover, a comparison
between lensing and rotation curve predictions can be done easily.

The virial theorem is a very useful tool to investigate the general
properties of the astrophysical systems. The scalar virial theorem gives a
very powerful constraint for the validity of the Thomas-Fermi approximation,
which can be formulated as
\begin{eqnarray}
R& \gg &1.581\times 10^{3} \left( \frac{m_{\chi }}{10^{-37}\;\mathrm{g}}%
\right) ^{1/2}\left( \frac{l_{a}}{10^{-20}\;\mathrm{cm}}\right) ^{-1/2}\times
\notag \\
&&\left( \frac{\rho _{c}}{10^{-24}\;\mathrm{g/cm^{3}}}\right)^{-1/2}\;
\mathrm{cm}.
\end{eqnarray}
The above constrain shows that the Thomas-Fermi approximation gives an
excellent description of the properties of condensate dark matter halos. By
using the tensor virial theorem we have derived the perturbation equation of
the dark matter halos, which can be efficiently used to study the stability
of dark matter halos under small perturbations.

Finally, we consider the problem of the mass of the dark matter particle.
The use of Eq.~(\ref{rad}) allows us to make a first estimate of the
physical properties of the dark matter particle. As a function of the mass
and scattering length of the particle the radius $R$ of the condensate dark
matter halo is given by $R=\pi \sqrt{\hbar ^{2}l_a/Gm_{\chi }^{3}}$, the
total mass of the condensate dark matter halo $M(R)$ can be obtained as
\begin{equation}
M(R)=4\pi ^2\left(\frac{\hbar ^2l_a}{Gm_{\chi }^3}\right)^{3/2}\rho _c=\frac{%
4}{\pi}\rho _{c}R^3 ,
\end{equation}
while the mean value $\left<\rho \right>$ of the condensate density is given
by the expression $\left<\rho \right>=3\rho _{c}/\pi ^2$ \citep{BoHa07}.
Therefore the dark matter particle mass in the condensate is given by %
\citep{BoHa07}
\begin{eqnarray}  \label{mass}
m_{\chi } &=&\left( \frac{\pi ^{2}\hbar ^{2}l_a}{GR^{2}}\right) ^{1/3}
\notag \\
& \approx & 6.73\times 10^{-2} \left[ l_a\left( \mathrm{fm}\right) \right]
^{1/3}\left[ R\;\mathrm{(kpc)}\right] ^{-2/3}\;\mathrm{eV}.
\end{eqnarray}
For $l_a\approx 1 $ fm and $R\approx 10$ kpc, the typical mass of the
condensate particle is of the order of $m_{\chi }\approx 14$ meV. For $%
l_a\approx 10^{6}$ fm, corresponding to the values of $l_a$ observed in
terrestrial laboratory experiments, $m_{\chi }\approx 1.44$ eV.

An important method of observationally obtaining the properties of dark
matter is the study of the collisions between clusters of galaxies, like the
bullet cluster (1E 0657-56) and the baby bullet (MACSJ0025-12). From these
studies one can obtain constraints on the physical properties of dark
matter, such as its interaction cross-section with baryonic matter, and the
dark matter-dark matter self-interaction cross section. If the ratio $\sigma
_m=\sigma /m_{\chi}$ of the self-interaction cross section $\sigma =4\pi
l_a^2$ and of the dark matter particle mass $m_{\chi }$ is known from
observations, with the use of Eq.~(\ref{mass}) the mass of the dark matter
particle in the Bose-Einstein condensate can be obtained as \citep{Har5}
\begin{equation}
m_{\chi }=\left(\frac{\pi ^{3/2}\hbar ^2}{2G}\frac{\sqrt{\sigma _m}}{R^2}%
\right)^{2/5}.
\end{equation}

By comparing results from X-ray, strong lensing, weak lensing, and optical
observations with numerical simulations of the merging galaxy cluster 1E
0657-56 (the Bullet cluster), an upper limit (68 \% confidence) for $\sigma
_m$ of the order of $\sigma _m<1.25\;\mathrm{cm^2/g}$ was obtained in \cite%
{Bul}. By adopting for $\sigma _m$ a value of $\sigma _m=1.25\;\mathrm{cm^2/g%
}$, we obtain for the mass of the dark matter particle an upper limit of the
order
\begin{eqnarray}
& m_{\chi } < 3.1933\times10^{-37}\left(\frac{R}{10\;\mathrm{kpc}}%
\right)^{-4/5} \left(\frac{\sigma _m}{1.25\;\mathrm{cm^2/g}}\right)^{1/5}\;%
\mathrm{g}  \notag \\
& = 0.1791\times\left(\frac{R}{10\;\mathrm{kpc}}\right)^{-4/5} \left(\frac{%
\sigma _m}{1.25\;\mathrm{cm^2/g}}\right)^{1/5}\;\mathrm{meV}.
\end{eqnarray}
This mass limit is consistent with the limit obtained from cosmological
considerations in \cite{Bo}. By using this value of the particle mass we can
estimate the scattering length $l_a$ as
\begin{eqnarray}
l_a < \sqrt{\frac{\sigma _m\times m_{\chi }}{4\pi }} = 1.7827\times 10^{-6}\;%
\mathrm{fm}.
\end{eqnarray}

A stronger constraint for $\sigma _m$ was proposed in \cite{Bul1}, so that $%
\sigma _m\in(0.00335\;\mathrm{cm^2/g},0.0559\;\mathrm{cm^2/g})$, giving a
dark matter particle mass of the order
\begin{eqnarray}
& m_{\chi }\approx \left(9.516\times 10^{-38}-1.670\times 10^{-37}\right)
\left(\frac{R}{10\;\mathrm{kpc}}\right)^{-4/5}\;\mathrm{g}  \notag \\
& = \left(0.053-0.093\right)\times \left(\frac{R}{10\;\mathrm{kpc}}%
\right)^{-4/5}\;\mathrm{meV},
\end{eqnarray}
and a scattering length of the order of
\begin{eqnarray}
l_a \approx \left(5.038-27.255\right)\times 10^{-8}\;\mathrm{fm}.
\end{eqnarray}
Therefore the galactic radii data and the Bullet Cluster constraints predict
a condensate dark particle mass of the order of $m_{\chi }\approx 0.1$ meV.
Recent results on the self-interacting dark matter cross section have been
presented in \cite{Bul2,Bul3}. By using collisions between galaxy clusters
as tests of the non-gravitational forces acting on dark matter, from the
dark matter's lack of deceleration in the bullet cluster collision one
obtains a self-interaction cross-section of the order of $\sigma_{DM}/m <
1.25$ cm$^2$/g (68\% confidence limit) for long-ranged forces \cite{Bul2}.
From the observation of 72 collisions a self-interaction cross-section $%
\sigma_{DM}/m < 0.47$ cm$^2$/g (95\% CL) was inferred.


\section*{Acknowledgements}

FSNL is supported by a Funda\c{c}\~{a}o para a Ci\^{e}ncia e Tecnologia
Investigador FCT Research contract, with reference IF/00859/2012, funded by
FCT/MCTES (Portugal), and acknowledges financial support of the Funda\c{c}%
\~{a}o para a Ci\^{e}ncia e Tecnologia through the grant
EXPL/FIS-AST/1608/2013 and UID/FIS/04434/2013.


\end{document}